\documentstyle[preprint,aps,eqsecnum,tighten]{revtex}

\newcommand{\beq}{\begin{equation}}
\newcommand{\eeq}{\end{equation}}
\newcommand{\beqa}{\begin{eqnarray}}
\newcommand{\eeqa}{\end{eqnarray}}

\newcommand{\bpks}{\mbox{$B_d \to \psi K_S$}}
\newcommand{\bdcpm}{\mbox{$B_d \to \Delta C= \pm 1$}}

\newcommand{\npks}{\mbox{$N_{\psi K_S}$}}

\newcommand{\ndcp}{\mbox{$N_{\Delta C=+ 1}$}}
\newcommand{\ndcm}{\mbox{$N_{\Delta C=- 1}$}}
\newcommand{\nddc}{\mbox{$N_{|\Delta C|=1}$}}
\newcommand{\nn}{\nonumber}
\newcommand{\Bb}{\bar{B}}

\newcommand{\Ima}{{\rm Im}\,}

\newcommand{\BBbar}{$B^0 - \bar B^0$}
\def\sm{Standard Model}
\def\BR{{\rm BR}}

\def\ie{{\it i.e.}}
\def\eg{{\it e.g.}}
\def\etal{{\it et al}}

\def\plb#1{Phys.\ Lett.\ {\bf B #1}}
\def\prd#1{Phys.\ Rev.\ {\bf D #1}}

\def\zpc#1{Z.~Phys.\ {\bf C #1}}

\draft
\begin{document}
{\tighten


\preprint{
\vbox{
      \hbox{SLAC-PUB-7834}
      \hbox{hep-ph/9805508}
      \hbox{May 1998} }}

\title{
Fast CP Violation\footnote{Research supported
by the Department of Energy under contract DE-AC03-76SF00515}}
\author{Yuval Grossman, Jos\'e R. Pel\'aez\footnote{On 
leave of absence from the Departamento de F\'{\i}sica Te\'orica.
Universidad Complutense. 28040 Madrid. Spain.}
and  Mihir P. Worah}
\address{ \vbox{\vskip 0.true cm}
Stanford Linear Accelerator Center \\
        Stanford University, Stanford, CA 94309}
\maketitle

\begin{abstract}%
$B$ flavor tagging will be extensively studied at the 
asymmetric $B$ factories due to its importance in 
CP asymmetry measurements. The primary tagging modes are the
semileptonic decays of the $b$ (lepton tag), or the 
hadronic $b \to c (\to s)$ decays (kaon tag).
We suggest that looking for time dependent 
CP asymmetries in events where one $B$
is tagged leptonically and the other one is tagged with a kaon
could result in an early detection of CP violation.
Although in the \sm\ these asymmetries are expected to be small,
$\sim 1\%$, they could be measured with about the same amount of data
as in the ``gold-plated'' decay \bpks.
In the presence of physics beyond the \sm, 
these asymmetries could be as large as $\sim 5\%$, and 
the first CP violation signal in the
$B$ system may show up in these events. 
We give explicit examples of new physics scenarios where this
occurs. 
\end{abstract}

}
\renewcommand{\thefootnote}{\alph{footnote}}

\section{Introduction}


One of the goals of the asymmetric $B$ factories is to study CP
violation in $B$ meson decays. CP violation has not been observed 
outside the kaon system,
thus it is important to identify $B$ decay modes that would allow an 
early detection of this phenomenon.  
The CP asymmetry in the \bpks\ decay is the benchmark to which all
other CP asymmetry measurements at the $B$ factories are usually compared 
\cite{review}. 
The branching ratio is relatively large ($5 \times 10^{-4}$) 
and the $\psi$ is easy to reconstruct from its
decay into two leptons. In addition, the CP asymmetry is expected to be
${\cal O}(1)$ in the \sm, and allow a clean measurement of 
the CKM angle $\beta$. 
It is also commonly assumed that this is the mode in which CP violation
will first be observed at the asymmetric $B$ factories. 

A crucial ingredient in the CP asymmetry measurements is flavor tagging.
In the asymmetric $B$ factories there are two main tagging techniques 
\cite{TDR}. The first is the ``lepton tag'', 
where the flavor is determined by the lepton charge
in a semileptonic $B$ decay. The second, the ``kaon tag'', uses 
hadronic $B_d$ decays to final states with $\Delta C =\pm 1$, namely
decays with one charmed hadron in the final state. These further
decay into a final state that contains only one
kaon, whose charge identifies the original
$B$ meson flavor.

In this work we propose that CP asymmetries in events where
both $B$'s are flavor tagged 
are also excellent candidates for an early observation of CP violation.
The cases where both $B$'s are tagged leptonically or both with kaons
result in CP asymmetries that are proportional only to
CP violation in the \BBbar\ mixing amplitude and, in particular, 
they vanish in the limit where the neutral $B$ width difference, 
$\Delta \Gamma$, is zero. The case we concentrate on here is 
where one of the $B$'s is tagged leptonically and the other using 
the kaon tag. This can lead to CP violation due to interference
between the neutral $B$ mixing and decay amplitudes.
Theoretically, these are given by the 
CP asymmetries in hadronic semi-inclusive
$B_d$ decays to final states with $\Delta C =\pm 1$.
Due to the importance of the kaon tag to the $B$ factories program,
the experimental issues
regarding the detection of these final states will be extensively
studied before and after the $B$ factories turn on. Moreover,
the search for these states should not result in much additional effort since
it will be carried out when performing the tagging.
Thus, we believe that the possibility of an early detection of CP
violation in these modes 
requires a simple extension of the flavor tagging studies.

The CP asymmetries we consider are the following:
\beq \label{dceqone}
a_{\Delta C=1}(t) \equiv
  {N_{B_\ell \bar B_K}(t) - N_{\bar B_\ell B_K}(t) \over 
  N_{B_\ell \bar B_K}(t) + N_{\bar B_\ell B_K}(t)},
\eeq
\beq \label{dceq-one}
a_{\Delta C=-1}(t) \equiv
{N_{B_\ell B_K}(t) - N_{\bar B_\ell \bar B_K}(t) \over 
  N_{B_\ell B_K}(t) + N_{\bar B_\ell \bar B_K}(t)},
\eeq
\beq \label{absdceqone}
a_{|\Delta C|=1}(t) \equiv
 {(N_{B_\ell B_K}+N_{ B_\ell\bar B_K})(t) 
- (N_{\bar B_\ell B_K} + N_{\bar B_\ell\bar B_K})(t) \over 
  (N_{B_\ell B_K}+N_{ B_\ell\bar B_K})(t) + 
(N_{\bar B_\ell B_K} +N_{ \bar B_\ell \bar B_K})(t)}.
\eeq
In the above notation $N_{B_\ell \bar B_K}$ is
the number of events where a $B$ has been tagged leptonically
and a $\bar B$ has been tagged with a kaon ($N_{B_\ell B_K}$ and 
$N_{\bar B_\ell \bar B_K}$ are similarly defined).
As we will show, 
in the \sm, the number of \BBbar\ pairs required
to get a statistically significant CP violation signal in \bdcpm\ is 
expected to be larger than that required in {\bpks}, yet achievable 
in approximately one year of data taking at nominal luminosity.
If, however, there is physics beyond the \sm, 
it could lead to the possibility of observing CP violation
in  \bdcpm\ in the very early stages of data taking, 
with up to an order of magnitude fewer \BBbar\ events than those
needed for the CP asymmetry in {\bpks}.

\section{Formalism and Standard Model Expectations}

The suggestion of looking for a CP asymmetry in the semi-inclusive
\bdcpm\ modes, the formalism to calculate it, as well as the \sm\
expectations were presented in a recent paper by Beneke,
Buchalla and Dunietz \cite{BBD}. Ignoring the width difference of the
neutral $B$ mesons ($\Gamma_{12}/M_{12} \simeq 0$), one obtains
\beq
a_{\Delta C=\pm 1}^{SM}(t) = {\cal C} \times
                             \frac{2\Ima\lambda_{SM} \sin\Delta mt}
{(1 \mp \cos\Delta mt)+|\lambda_{SM}|^2(1 \pm \cos\Delta mt)},
\label{acp_pm}
\eeq
and
\beq
a_{|\Delta C|=1}^{SM}(t) \simeq {\cal C} \times
                             2{\Ima\lambda_{SM} \sin\Delta mt},
\label{acp_abs}
\eeq
where $\lambda_{SM} =
e^{i2\beta}(V_{ub}^{*}V_{cd}^{})/(V_{cb}^{*}V_{ud}^{})$ and ${\cal C}$ is a
dilution factor that arises because one is summing over several
exclusive modes, possibly with opposite CP asymmetries.
Assuming local quark-hadron 
duality, setting all bag factors equal to one, and using 
$f_B = 180$ MeV, $m_b=4.8$ GeV and $m_c=1.4$ GeV, one finds 
${\cal C} = -0.21$ \cite{BBD}. 
Using $|V_{ub}|/|V_{cb}|=0.1$ one then obtains 
\beq
a_{\Delta C=\pm 1}^{SM}(t) \simeq \frac
                   {-0.01\sin(\alpha - \beta)\sin\Delta mt}
{(1 \mp \cos\Delta mt)+0.0005(1 \pm \cos\Delta mt)}
\label{adcpm_sm_est}
\eeq
and
\beq
a_{|\Delta C|=1}^{SM}(t) \simeq -0.01 \;
                   {\sin(\alpha - \beta)\sin\Delta mt}.
\label{addc_sm_est}
\eeq
For comparison, the CP asymmetry in \bpks\ is given by
\beq
a_{\psi K_S}^{SM}(t) = \sin(2\beta)\sin\Delta mt,
\label{apks_sm}
\eeq
with essentially no uncertainty.
We consider the value of ${\cal C}$ used above only 
as a reasonable estimate due to the large uncertainties
in some of the factors that go into its determination. 
Until these are better known, the CP asymmetries in 
Eqs. (\ref{adcpm_sm_est}) and (\ref{addc_sm_est}) are unlikely 
to be of much use in
obtaining precise measurements of CKM angles, as suggested in
\cite{BBD}. The importance of these modes lies, rather, in
the fact that they may lead to an early detection of CP violation;
possibly the first at the $B$ factories.

The number of \BBbar\ events needed to establish CP violation
depends not only on the expected CP asymmetry, but
also on the $B$ meson branching ratio as well as on tagging and detection
efficiencies for that particular final state.
Thus, the small CP asymmetry in \bdcpm\ is compensated for by the large
branching ratio into these modes. 
We can estimate the number of \BBbar\ pairs ($N_f$)
required to obtain a $3\sigma$ CP violation signal in a given final
state $f$ using
\beq
N_{f}  \epsilon_f \BR_f \gtrsim \frac{10}{a_{f}^2},
\label{num_1}
\eeq
where $\epsilon_f$ is the combination of detection and $B$ flavor
tagging efficiencies, $\BR_f$ the branching ratio, and $a_f$ is 
a time integrated CP asymmetry for the final state $f$. 
We present below the results based on Eq. (\ref{num_1}) for the
various modes, relegating details concerning its implementation 
to the Appendix.

For the $\Delta C = \pm 1$ modes we find in the \sm,
\beqa
\nddc &\gtrsim&  
\displaystyle{{2 \times 10^{7}  \over
\sin^2(\beta-\alpha)}}, \\
\ndcp &\gtrsim& 
\displaystyle{{2 \times 10^{7}  \over 
\sin^2(\beta-\alpha)}}, \nonumber \\
\ndcm &\gtrsim& 
\displaystyle{{6 \times 10^{7}  \over 
\sin^2(\beta-\alpha)}} \nonumber .
\eeqa
For comparison, the number of \BBbar\ pairs 
needed to establish a CP violation
in \bpks\ can be estimated to be about 
\beq 
\npks\gtrsim {3 \times 10^{6} \over \sin^2 (2\beta)}
\eeq
Thus, one can expect to observe CP violation in the \bdcpm\ modes with 
roughly an order of magnitude more data than that needed 
in the ``gold-plated'' \bpks\ mode. This amount of data will be available 
after one to two years of data taking with the expected luminosity
yielding $\sim 10^7$ \BBbar\ pairs per year.
If it turns out that 
$|{\cal C}| > 0.21$ or if $|\sin(\alpha-\beta)|>|\sin(2\beta)|$,
it may be possible to observe CP violation in the \bdcpm\ modes with a
comparable amount of data as in \bpks\ even in the \sm.
Given the fact that, in any case, data will be
taken and analyzed in the \bdcpm\ modes, it is important then  
to also search for CP violation.

\section{New Physics}

The situation gets even more interesting in the presence of physics
beyond the \sm. 
Note from Eq. (\ref{num_1}) that $N_f$ scales as $(a_{f})^{-2}$,
and that the small value of $(a_{\Delta C=\pm 1})$ in
Eqs. (\ref{adcpm_sm_est}) and (\ref{addc_sm_est}) is 
essentially due to the small ratio of amplitudes,
\beq
\left|{A(b \to u \bar c d) \over A(b \to c\bar u d)}\right|_{SM} = 
\lambda \left|{V_{ub} \over V_{cb}}\right|.
\eeq 
However, the $b \to u \bar c d$
rate is not well measured at present and could contain large CP violating
new physics contributions. If such new contributions significantly enhance 
the magnitude of that amplitude, they will also enhance 
the CP asymmetries in these modes.
 
Let us first present a model independent analysis of the
bounds on physics beyond the \sm\ for the $b \to u \bar c d$
transition. 
We parameterize the new physics effects by the quantity 
\beq
r \equiv {A(b\to u \bar c d)_{NP} \over A(b\to u \bar c d)_{SM}}.
\label{def_r}
\eeq
In the limit $\vert r\vert \gg 1$, Eqs. (\ref{adcpm_sm_est}) and
(\ref{addc_sm_est}) become%
\footnote{%
We ignore new physics contributions to \BBbar\ mixing 
and $b\to c \bar u d$ decay since they cannot significantly enhance
the CP asymmetry in \bdcpm. See Refs. \cite{GNW} and \cite{GW},
respectively, for a discussion on detecting 
such contributions using CP violating $B$ decays.}
\beq
a_{\Delta C= -1}(t) \simeq  \frac
       {- \vert r\vert \; 0.01\sin(2 \beta + \theta)\sin\Delta mt
-2a_{SL}\sin^2({\Delta mt}/{2})}
{(1 - \cos\Delta mt)+0.0005\vert r\vert ^2(1 + \cos\Delta mt)},
\label{adcm_np}
\eeq
\beq
a_{\Delta C= +1}(t) \simeq  \frac
       {- \vert r\vert \; 0.01\sin(2 \beta + \theta)\sin\Delta mt
-0.001a_{SL}\sin^2({\Delta mt}/{2})}
{(1 + \cos\Delta mt)+0.0005\vert r\vert ^2(1 - \cos\Delta mt)},
\label{adcp_np}
\eeq
and
\beq
a_{|\Delta C|=1}(t) \simeq -0.01 \vert r\vert
 \sin(2\beta+\theta)\sin\Delta mt
-2a_{SL}\sin^2({\Delta mt}/{2}),
\label{addc_np}
\eeq
respectively, 
where $\theta$ is the phase between $A(b\to u \bar c d)_{NP}$ and 
$A(b\to c \bar u d)_{SM}$. Moreover, we have included the term
proportional to $a_{SL} \equiv {\Ima}(\Gamma_{12}/M_{12})$ \cite{BBD}
which is a measure
of the CP violation in events with two lepton tags, \ie
\beq
\frac{N_{B_lB_l}(t)-N_{\bar B_l\bar B_l}(t)}
     {N_{B_lB_l}(t)+N_{\bar B_l\bar B_l}(t)}=
      a_{SL} \sin^2\frac{\Delta mt}{2}.
\eeq
and we are using the same notation as in
Eqs. (\ref{dceqone}-\ref{absdceqone}). This is expected to be
negligible in the \sm, but is likely to get large enhancements from a
new $b \to u \bar c d$ amplitude.  
Thus, we see that for $r \sim 5$ the CP asymmetry can be enhanced
by a factor of 5, and hence a CP violation signal can be obtained with  a
factor 25 less data than required in the \sm.
Note that since we care only about order of magnitude
estimates, we have ignored corrections associated with
generalizing the results of \cite{BBD} to account for operators with
non-standard Lorentz structures. 

We now present constraints on the new physics contributions if it consists of
right-handed currents contributing to the $b\to u\bar c d$ decay.
There are
no reported bounds on either the inclusive or exclusive decays. 
In order to derive such a constraint, we require 
that the branching ratio for this mode
should not be large enough to have a significant effect on the $B$
semi-leptonic branching ratio. This implies 
$\BR(b \to u \bar c d) \lesssim 10\%$. 
Comparing to the \sm\ expectation
$\BR(b \to u \bar c d)_{SM} \sim 1 \times 10^{-4}$ we obtain 
that $r \lesssim 30$.\footnote{%
Note that such a large new contribution to the $b \to u \bar c d$
decay could even help reconcile the differences between the
theoretical prediction and the results of the $B$
semi-leptonic branching ratio and charm counting  experiments
\cite{russians-alex}.} 

Considerations of exclusive decays also lead to similar bounds. For
example, $r \sim 10$ combined with 
$\BR(B^+ \to \bar D^0 \rho^+) \sim 0.5 \%$ leads to   
$\BR(B^+ \to D^0 \pi^+) \simeq 1.2 \times 10^{-4}$. Although CLEO has no
bounds on this process, it is likely that a dedicated search could obtain a
bound $\BR(B^+ \to D^0 \pi^+)\sim  10^{-4}$.
A larger enhancement up to $\BR(B^+ \to D^0 \pi^+)\sim  10^{-3}$
is unlikely, as this might result in intrinsic inconsistencies in other
measurements such as $B \to DK$ \cite{Avner}. 

Finally, this new amplitude could result in a large enhancement of the
semi leptonic CP asymmetry $a_{SL}$ essentially because it makes a
large contribution
to final states common to both $B^0$ and $\bar B^0$
mesons, and moreover the GIM mechanism which further suppresses
the \sm\
result is no longer effective. If this new physics effect dominates
$\Gamma_{12}$ one can derive the bound
$a_{SL} \sim 0.015|r|$. 
Using OPAL's $2\sigma$ limit, $a_{SL} \lesssim 0.08$ \cite{OPAL}, 
one obtains the constraint $r \lesssim 5$.
Note, however, that this bound could be modified by a
factor of a few if there are new contributions to $M_{12}$.

As a simple example of a new physics scenario that can lead to such
large values of $r$, and thus enhanced CP violation in \bdcpm, we
consider a non minimal Left-Right Symmetric Model \cite{tom}. 
Namely, one where
the left ($V^L$) and right ($V^R$) quark mixing matrices are not
related. We assume identical gauge couplings, $g_L=g_R$,
and that the $W_L-W_R$ mixing is negligible. 
Then, tree level $W_R$ 
exchange can lead to the desired final state. The ratio of the new amplitude
to the \sm\ one is 
\beq
|r| \approx \left |V_{ub}^R V_{cd}^R \over V_{ub}^L V_{cd}^L\right| 
\left( m_{W_L}^2 \over m_{W_R}^2 \right).
\eeq
For $|V_{ub}^R V_{cd}^R| \simeq 1/2$ and $m_{W_R} \simeq  10\,m_{W_L}$ we get 
$r \simeq 5$, while still satisfying all other constraints on the
model. Of course, we also assume that the CP violating phase in $V^R$ is 
large, $O(1)$.

In this case the CP asymmetries in the various tagging modes are enhanced 
and can reach the $5\%$ level. 
As a consequence, we would observe a CP violation signal
in events with one lepton and one kaon tag, as well as
in events with both lepton or both kaon tags with
about $10^6$ \BBbar\ pairs. 
Moreover, such enhanced CP violation would be a clear signal of 
physics beyond the \sm. Of course, if no such signal is found, we will be able
to put bounds on the magnitude of new contributions to the $b \to u \bar c d$
amplitude.

We have also studied other scenarios, \eg, models with extra
charged scalars, models with diquarks, supersymmetry with four
generations and broken $R$ parity, all of which allow $r
\sim 5$, and hence the possibility of an early detection of CP violation. 

\section{Conclusions}

We have proposed that it is important to search for CP
violation in events where one $B$ has been tagged leptonically, and
the other $B$ by a kaon. Within the \sm, the number of \BBbar\
events required to detect 
CP violation in this mode could be similar to that for \bpks, and could be
obtained in the first year of running at the $B$ factories.
In the presence of new physics, it is 
possible that CP violation could be detected  in \bdcpm\ with 
significantly less data than needed to detect it in \bpks, and thus be
the first CP violating signal at the asymmetric $B$ factories.
Moreover, such new physics is likely to contribute to the neutral $B$ width
difference, resulting in observable CP violation also in the modes
where both $B$'s are tagged either leptonically, or with kaons. 
Given the fact that $B$ flavor tagging will be intensively 
studied, we suggest that the possibility of observing CP violation in 
these modes should be seriously considered. 

\acknowledgments

The authors thank Y. Nir,
S. Plaszczynski, H. Quinn, T. Rizzo, J. Rosner, M-H. Schune, A. Snyder and
A. Soffer for
useful conversations. J. R. P. has been partially supported by the
Spanish CICYT under contract AEN93-0776, and 
thanks the SLAC Theory Group 
for their kind hospitality as well as the Spanish 
Ministerio de Educaci\'on y Cultura for a fellowship.


\appendix
\section{Numerical Estimates}

The number of \BBbar\ pairs required to obtain a
statistically significant observation of CP violation has been estimated 
in the text by means of  
\beq
N_{f}  \epsilon_f \BR_f \gtrsim \frac{10}{a_{f}^2}
\eeq
where, 
given a time interval $(a,b)$, we have defined the CP asymmetry
\beq
a_{f}=\frac{\int_a^b {\cal N}_f(t)\,dt}{\int_a^b {\cal D}_f(t)\,dt},
\eeq
with
\beqa
{\cal N}_f(t)&=&\Gamma(B(t)\rightarrow f)-\Gamma(\Bb(t)
                  \rightarrow \bar{f}), \nn\\
{\cal D}_f(t)&=&\Gamma(B(t)\rightarrow f)+\Gamma(\Bb(t)
                  \rightarrow \bar{f}).
\eeqa
Moreover, the number of neutral $B$ decays are an oscillatory function
of time, and hence we use 
\beq
\BR_f=\frac{\int_a^b {\cal D}_f(t)\,dt}{2\Gamma}
\eeq
where $\Gamma$ is the total $B$ width, finally leading to
\beq
N_{f} \gtrsim \frac{20\,\Gamma}{\epsilon_f}
            \frac{\int_a^b {\cal D}_f(t)\,dt}
                 {\left(\int_a^b {\cal N}_f(t)\,dt\right)^2}.
\label{num_2}
\eeq
Here we are assuming that we have already reconstructed the time information
of the decay, since otherwise we could not make the $\int_a^b$ integral.
Although in practise the CP violating signal is best extracted 
by using all the data to perform a 
maximal likelihood fit to Eqs. (\ref{acp_pm}) and (\ref{acp_abs}),
or their generalizations in the case of physics beyond the \sm,
we believe that the above equation yields a 
reasonable estimate of the number of \BBbar\ pairs required 
to see a CP violating signal.

Using the
information in the BABAR Technical Design Report, and 
recent studies made for the BABAR physics book \cite{TDR}, we 
estimate $\epsilon_{\Delta C=1} \approx 2.5\times 10^{-2}$ for all the
$\Delta C=1$ modes. This estimate is a product of 10\% for the lepton
tag, 50\% for the kaon tag and a further 50\% due to the fact that in
the inclusive decay one needs to identify both the flavor and the
charge of the decaying $B$. Similarly, we estimate 
$\epsilon_{\psi K_S}\approx 1.5 \times 10^{-2}$ as a product of 30\% for the 
combined
lepton and kaon tags, 10\% for the $\psi$ reconstruction, and 50\% 
for the $K_S$ reconstruction. We have ignored other possible
systematic differences in observing the CP asymmetries in the two
modes such as the fact that the vertexing efficiencies may be
different in the two cases, or that the purity required of the kaon
tag may be different. We deem these to be subjects for a more detailed
analysis than undertaken here.

When doing the integral in Eq. (\ref{num_2}) we utilize a common lower
limit of 0.6 ps corresponding to a resolvable separation of 100 $\mu$ for
the two $B$'s in the BABAR environment, 
and optimize the upper bound to get the best CP
signal. The $\Delta C=+1$ mode is particularly sensitive to the
vertexing resolution since the
smallness of the $D(t)$ at early times [cf Eq. (\ref{acp_pm})]
combined with the lack of
exponential suppression from the $B$ lifetime, implies that 
one could significantly enhance the CP signal by 
giving more weight to the early time region.

\nopagebreak
{\tighten

}


\begin{thebibliography}{ }

\bibitem{review}
For a review see e.g.,
Y. Nir and H.R. Quinn, Ann. Rev. Nucl. Part. Sci. {\bf 42} 211 (1992);
Y. Nir, Lectures presented in the 20th SLAC Summer Institute,
SLAC-PUB-5874 (1992).

\bibitem{TDR}
BABAR Technical Design Report, SLAC-R-95-457;
The BABAR Physics Book, SLAC-R-504, in preparation.

\bibitem{BBD}
M. Beneke, G. Buchalla and I. Dunietz, \plb{393} 132 (1997).

\bibitem{GNW}
Y. Grossman, Y. Nir and M. Worah, \plb{407} 307 (1997).

\bibitem{GW}
Y. Grossman and M. Worah, \plb{395} 241 (1997).

\bibitem{russians-alex}
For an introduction and possible resolutions see e.g.,
I. Bigi {\em etal}, \plb{323} 408, (1993); A. Kagan, \prd{51}, 6196 (1995).

\bibitem{Avner}
A. Soffer, private communication. 

\bibitem{OPAL}
The OPAL collaboration, K. Ackerstaff \etal, \zpc{76}, 401 (1997).

\bibitem{tom}
For a recent analysis see e.g., T. Rizzo, hep-ph/9803385, and references 
therein.

\end{thebibliography}
\end{document}